%% file: DPF_Fiedler.tex
\documentclass[12pt]{article}
\usepackage{epsfig}
\usepackage{amsmath}
\usepackage{float}
\usepackage{color}

\input econfmacros.tex
\def\Title#1{\begin{center} {\Large {\bf #1} } \end{center}}

\begin{document}
\renewcommand{\thefootnote}{\fnsymbol{footnote}}

\Title{Investigation of Beam Emittance and Beam Transport Line Optics on Polarization}

\bigskip\bigskip


\begin{raggedright}  

{\it Andrew Fiedler$^{1}$ and Michael Syphers$^{1,2}$ \index{Fiedler, A.}\\
$^{1}$Department of Physics, 
Northern Illinois University\\
 $^{2}$Fermi National Accelerator Laboratory\footnote[1]{Operated by Fermi Research Alliance, LLC under Contract No. DE-AC02-07CH11359 with the United States Department of Energy.
} \\
Poster presented at the APS Division of Particles and Fields Meeting (DPF 2017), July 31-August 4, 2017, Fermilab. C170731}
\bigskip\bigskip
\end{raggedright}

\section{Introduction}
The Muon $g$-2 Experiment \cite{TDR}, currently in its commissioning phase at Fermi National Accelerator Laboratory (Fermilab), aims to measure the value of the anomalous magnetic moment of the muon to unprecedented accuracy.   This experiment and other similar experiments involving precessions due to magnetic dipole moments and possibly electric dipole moments, rely on a highly polarized beam for success.  Hence, understanding the effects of phase space variables and field perturbations on polarization is necessary for these high precision experiments. In this work, the Muon $g$-2 beam delivery system at Fermilab is used as an example and the results of computer simulations of the effects of beam optics on polarization are examined.

To begin, we note that the time-dependent behavior of particle spin under the influence of electric and magnetic fields can be described by the Thomas-BMT equation \cite{Thomas}, \cite{BMT}: 

\begin{equation}
\frac{d \vec{S}}{dt} = \frac{e}{\gamma_r m} \vec{S} \times \left[ (1 + a \gamma_r) \vec{B}_{\perp} + (1 + a) \vec{B}_{\parallel} + \left(a \gamma_r + \frac{\gamma_r}{\gamma_r + 1} \right) \frac{\vec{E} \times \vec{\beta}}{c} \right]
\label{eq:ThomasBMT}
\end{equation}
where $a$ is the anomalous magnetic moment, defined as
\begin{equation}
a = \frac{g-2}{2}
\end{equation}
and the values of the fields are those in the laboratory frame, whereas the spin vector $\vec{S}$ is in the rest frame of the particle.  Here, $g$ is the gyromagnetic factor which, for a pure Dirac particle, would have value of $g=2$.

If we examine the special case where there is no electric field and the magnetic field is perpendicular to the motion of the charged particle, we can reduce Eq.~\ref{eq:ThomasBMT} to
\begin{equation}
\frac{d \vec{S}}{dt} = \frac{e}{\gamma_r m} (1 + a \gamma_r) \vec{S} \times \vec{B}_{\perp}. \label{e:bmt}
\end{equation}
In a high energy beam transport system composed of magnetic elements with predominantly transverse guiding and focusing fields, we can use this result to make approximations of effects on beam polarization due to several common realistic beam characteristics such as transverse beam emittance, momentum spread, and error fields due to magnet misalignments and mispowerings.  These can be compared to the natural polarization created in the manifestation of muons from pion decay in the production of the beam to be used in the experiment.

\section{Effects on Polarization}

Using a classical description as presented in Eq.~\ref{e:bmt}, one can imagine the spin of a particle being aligned in a particular direction in space and the degree of polarization being the degree to which the spins of an ensemble of particles are aligned with each other.  With this picture in mind we will examine below three particular properties that can affect the spread in spin directions of the particles and finally we compare these effects to the polarization, or spread in spin directions of the muon beam produced from pion decay. 

\begin{figure}[htb]
\begin{center}
\epsfig{file=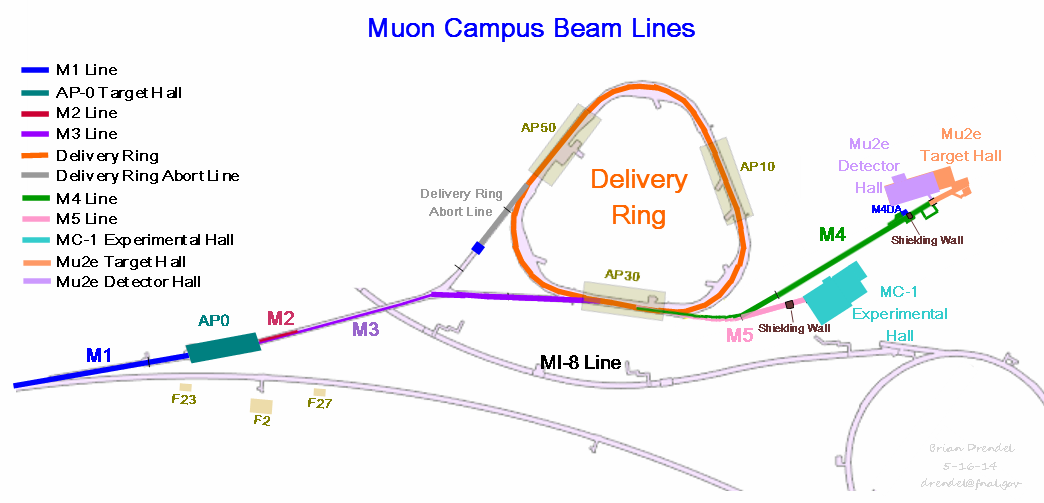,height=2.5in}
\caption{Diagram of the beam lines for muon delivery system at Fermilab. Courtesy Brian Drendel.}
\label{fig:Beamlines}
\end{center}
\end{figure}

While the disussions below will generate analytical estimates of the magnitudes of these effects, computer simulations of particle transport along the muon delivery system for the Muon $g$-2 experiment were also performed, using G4beamline \cite{Roberts}.  This code can include particle decays as well as keep track of particle polarization. The simulation used a distribution of particles (protons, muons and pions) produced from the targeting 8 GeV (kinetic energy) protons onto the Fermilab AP0 target, and then transported them through the M2 and M3 beam lines to the Delivery Ring, where they circulate 4 times, allowing the pions to decay to muons and the muons to separate from protons by time-of-flight. The resulting muon beam is then extracted and transported to the experiment (in MC-1) via the M4 and M5 lines.   A schematic of this roughly 3~km-long system is provided in Fig.~\ \ref{fig:Beamlines}.

Below, we will examine how momentum spread, beam emittance, and magnet misalignments contribute to the depolarization of a beam and estimate the magnitude of the contribution of each effect on the beam's polarization.  Comparisons will then be made with the polarization generated during the creation of the muons from pion decay.

\subsection{Momentum Spread}
From Eq.~\ref{eq:ThomasBMT} we can see that the spin vector of a particle passing through a bending magnet of length $\ell$ and field strength $B$ will precess through an angle in the bending plane of amount
\begin{equation}
\Delta \phi = \frac{d\vec{S}}{dt}\cdot\frac{\ell}{v} = \frac{eB\ell}{\gamma_r mv} \cdot (1+a \gamma_r) = (1+a\gamma_r)\theta
\label{eq:momentum}
\end{equation}
where $a$ is the anomalous magnetic moment, $\gamma_r$ is the relativistic gamma factor, and $\theta$ is the bending angle of the magnet.  We assume here small precession angles anticipating that $a$ and $\theta$ are small quantities (note:  for the muon, $a \approx 10^{-3}$).  Since the central trajectory of the particle beam bends through an angle $\theta$, then relative to this ideal trajectory the {\em additional} rotation of the spin vector is $\Delta\phi = a\gamma_r\theta$.  From now on we will only keep track of these additional rotations with respect to the central trajectory in our analyses.\\

The constituents of a particle beam will have a small range of momenta and hence they will be bent by differing amounts through a bending element as also will their spins. A particle beam with a spread in momenta $\Delta p/p$ will thus have a spread in spin rotation angles upon passing through the bending magnet given by
\begin{equation}
\Delta \phi_{rms} = a \gamma_r \theta \left( \frac{d \gamma_r}{\gamma_r} \right)_{rms}
 = a \gamma_r \theta \left( \frac{\Delta p}{p} \right)_{rms}
\end{equation}
where, for the case of a highly relativistic beam, $d \gamma_r / \gamma_r~\approx~dp / p$. In the case of the $g$-2 M2M3 beamline, we can take $a \approx 0.001$, $\gamma_r \approx 30$, and $\frac{dp}{p} \approx 1.5 \% $. 
The total bending angle of the beam line is $\theta \approx 10^{\circ}$, which corresponds to 0.2 rad. This allows us to estimate that the rms spread in spin directions will be characterized by $\Delta \phi_{rms} \approx 0.1$~mrad upon passage through these beam lines.  Performing the same estimation with the Delivery Ring (changing the value of $\theta$ to $8 \pi$ radians), gives $\Delta \phi_{rms} \approx 11$ mrad. 

\subsection{Emittance}
The size of the beam is another source of depolarizing effects, due to the field strength of focusing quadrupoles varying with distance from the ideal path.  A larger beam contains particles further from the center, and thus, those particles experience a stronger correcting field and greater precession. 
The general solution for the transverse position of a particle traveling along a beamline is given in terms of the amplitude function, or $\beta$ function, according to
\begin{equation}
x (s) = A \sqrt{\beta (s)} \sin (\psi + \delta)
\label{eq:gensolution}
\end{equation}
where $\psi(s)$ is the phase advance, related to $\beta(s)$ by $d\psi(s)/ds = 1/\beta(s)$.  

\begin{figure}
\centering
\includegraphics[width=0.4\linewidth]{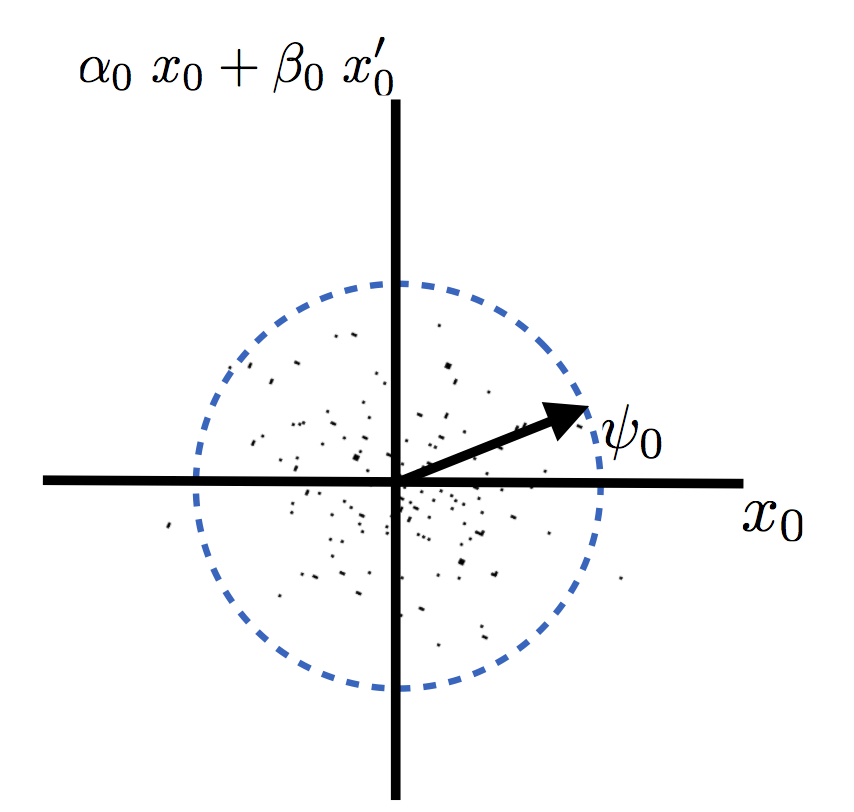}
~~~~~~~
\includegraphics[width=0.4\textwidth]{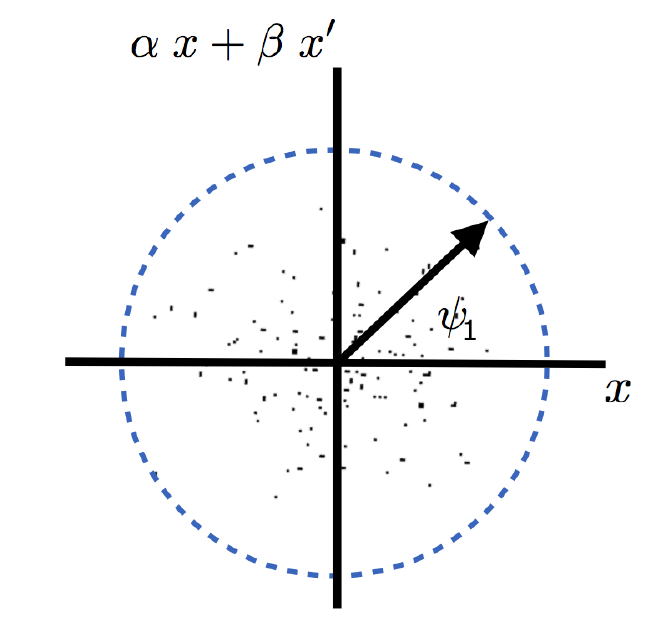}
\caption[Emittance of circular distribution]{The figures above denote the changes in phase for a distribution under the transformation $y=\alpha x + \beta x'$. Under this transformation, the circular distribution rotates (noted by the change in the angle $\psi$) as the beam traverses the lattice.}
\label{fig:emittance_circles}
\end{figure}

It is common to describe the beam ensemble in terms of the particle's coordinates $x$ and $x' = dx/ds$ in $x-x'$ phase space.  The elliptical paths in this phase space as the particles travel down the beam line can be reduced to circular phase space trajectories by the transformation
\begin{align}
x &= x \\ 
y &= \beta x' + \alpha x
\end{align} 
with $\alpha \equiv -\frac12 d\beta/ds$.  In our new phase space coordinates the motion can be described as a rotation along with an appropriate scaling factor given by the values of $\beta$ at the starting and ending points as generated by the optical system. 
This is illustrated in Fig.~\ref{fig:emittance_circles}, and expressed by 
\begin{equation}
\left( \begin{array}{c}
x \\ y
\end{array} \right)
= \sqrt{\frac{\beta}{\beta_0}} 
\left( \begin{array}{cc}
\cos \psi & \sin \psi \\
- \sin \psi & \cos \psi
\end{array} \right)
\left( \begin{array}{c}
x_0 \\ y_0
\end{array} \right)
\label{eq:circemit}
\end{equation}
where $\psi$ represents the phase advance between the two end points.

As a particle oscillates through the focusing system its angle $x'$ with respect to the ideal trajectory will change in accordance to Eq.~\ref{eq:circemit} and hence the total change in angle in going from location $s_0$ to location $s$ will be $\Delta x' = x'(s) - x'(s_0)$.  Using Eq.~\ref{eq:momentum} we see that the spin direction of the particle will rotate through an angle
\begin{equation}
\Delta \phi = (1+a \gamma_r) \Delta x'.
\end{equation}
Note that the reference trajectory is not altered by the purely focusing elements, and so the leading factor in this last equation is $1+a\gamma_r$ and not $a\gamma_r$.

Using our new transformation to solve for $x'$ in terms of initial conditions $x_0$ and $y_0 = \beta_0 x'_0 + \alpha_0 x_0$, we find that
\begin{equation}
x' = \frac{1}{\sqrt{\beta_0 \beta}} \left[ -(\sin \psi + \alpha \cos \psi) x_0 + (\cos \psi - \alpha \sin \psi) y_0  \right].
\end{equation}
So, in going from location $s_0$ to location $s$, the angle of the particle's trajectory relative to the design trajectory will change by an amount
\begin{equation}
\Delta x' = x' - x_0' = \frac{1}{\sqrt{\beta_0 \beta}} \left[ - (\sin \psi + \alpha \cos \psi - r \alpha_0 ) x_0 + (\cos \psi - \alpha \sin \psi - r) y_0 \right]
\end{equation}
where $r \equiv \sqrt{\beta / \beta_0}$.  By squaring this result and averaging over the ensemble of particles it follows that
\begin{multline}
\Delta x_{rms}' = \sqrt{\frac{<x_0^2>}{\beta_0} \frac{1}{\beta} \left[ 1 + \alpha^2 - 2r(1 + \alpha_0 \alpha) \cos \psi + r^2 (1 + \alpha_0^2) + 2r(\alpha - \alpha_0) \sin \psi \right]}
\end{multline}
where we note that $\langle x^2 \rangle = \langle y^2 \rangle$ in our chosen cylindrically symmetric coordinate system. 

We can immediately identify the quantity $\pi \langle x_0^2\rangle/\beta_0$ as the transverse emittance of the beam in the $x$ degree of freedom.  In addition, we can also identify the quantity $(1+\alpha^2)/\beta \equiv \gamma$; the combination of $\beta$, $\alpha$, and $\gamma$ as defined above are known as the Courant-Snyder parameters.\cite{Courant}  Thus, the rms value of the changes in particle spin angles when transported between two points in our beam lines can be written as
\begin{equation}
\Delta \phi_{rms} = (1 + a_{\mu} \gamma_r) \sqrt{\frac{2 \epsilon_{rms}}{\pi}} \sqrt{\left[  \frac{\gamma + \gamma_0}{2} - \frac{1 + \alpha_0 \alpha}{\sqrt{\beta_0 \beta}} \cos \psi + \frac{\alpha - \alpha_0}{\sqrt{\beta_0 \beta}} \sin \psi \right]}.
\label{e:rmsSpin}
\end{equation}

To summarize, a particle transported between locations $s_0$ and $s$ will have its spin direction altered by an amount $\Delta \phi = \phi-\phi_0$.  To the extent that the change in $\phi$ is independent of its initial orientation, then Eq.~\ref{e:rmsSpin} shows the extent to which the beam will depolarize due to the fact that the particles with larger betatron amplitudes will be steered more strongly by the focusing quadrupoles and hence will have larger precessions.  Note that if the transport system begins and ends with identical Courant-Snyder parameters, then our expression reduces to 
\begin{equation}
\Delta\phi_{rms} = 2(1+a_\mu\gamma_r)\sqrt{\epsilon_{rms}\gamma_0/\pi} \;\sin\psi
\end{equation}
and if the phase advance through this transport system is $\psi = 2\pi$, then the polarization of the beam returns to its initial value.

For our M2M3 delivery line, we can use Eq.~\ref{e:rmsSpin} to estimate the rms spin spread based on the input parameters, $a_\mu \approx 0.001$, $\gamma \approx 30$, $\beta_0 = 2.488~$m, $\beta = 5.0328~$m, $\alpha_0~=~0.175$, $\alpha$ = -0.72738 and an rms emittance of 7$\pi$ mm$\cdot$mrad (assuming a $40\pi$ (95\%) emittance for a Gaussian beam (\cite{TDR} p.200)) gives $\Delta \phi_{rms} \approx 2.7$~mrad. Statistical analysis of a sample of 500 muons (with decays turned off) through the M2M3 delivery line gave a value of $\Delta \phi_{rms} \approx 2.8$, which is close to our numeric approximation.

\subsection{Misalignments}
Similar to emittance effects, there are consequences to the rms spread of spin due to magnetic misalignment.   Analytically, we can treat each misalignment separately and sum their effects (for a first-order approximation). This allows us to write the change in slope, $\Delta x'$ as the ratio of the displacement of the magnet to its focal length. 
\begin{equation}
\Delta x' = \frac{d}{F}
\end{equation}
Summing displacements along a lattice gives the result
\begin{equation}
\Delta x' = \sum_{i=1}^N \frac{d_i}{F_i}(\cos \psi_i - \alpha_f \sin \psi_i) \sqrt{\frac{\beta_i}{\beta_f}}
\end{equation}
where $\psi_i$ is the phase advance from the $i-th$ misaligned element to the end of the beam transport system.  By squaring and summing terms, it follows that our random displacements would generate an rms change in transverse angle given approximately by
\begin{equation}
\Delta x_{rms}' = d_{rms} \sqrt{ \left< \frac{\beta}{F^2} \right> \frac{1}{\beta_f} \left( \frac{1}{2} + \frac{\alpha_f^2}{2} \right)} \sqrt{N} = d_{rms} \sqrt{\gamma_f \left< \frac{\beta}{F^2} \right>} \sqrt{\frac{N}{2}}
\end{equation}
Thus, the spread in the change of spin directions is estimated to be
\begin{equation}
\Delta \phi_{rms} = (1 + a \gamma_r) d_{rms} \sqrt{\gamma_f \left< \frac{\beta}{F^2} \right>} \sqrt{\frac{N}{2}}
\end{equation}
Once more using $a \approx 0.001$, $\gamma_r \approx 30$ and the known values for the M2M3 beamline ($\beta = 5~$m and  $\alpha = -0.73$), and using an rms displacement of 0.25 mm, and typical M2M3 values for $\frac{\beta}{F^2} = 1.3~\mathrm{m}^{-1}$ for 60 quadrupoles gives a $\Delta \phi_{rms} \approx 1$ mrad.

\subsection{Particle Decays}
\begin{figure}[htb]
\begin{center}
\epsfig{file=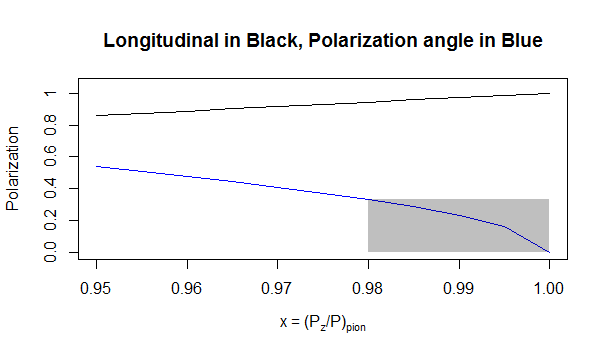,height=3.5in}
\caption{The graph shows the relationship between the momentum ratio of pions and the beam polarization. The black line represents the longitudinal beam polarization as found in Eq.~\ref{e:Combley}, while the blue line represents the corresponding spin angle (in radians).}
\label{fig:phi_decays}
\end{center}
\end{figure}

When pions decay into muons, a preference exists for the spin of the muon to be aligned with the momentum vector \cite{Garwin}.  Combley and Picasso \cite{Combley} have shown that in terms of the momentum ratio $x = p_\parallel / p_\pi$, where $p_\parallel$ is the component of the muon's momentum that is in the direction of the decayed pion, the resulting  longitudinal polarization ($\Sigma_L$) can be described by the equation
\begin{equation}
\Sigma_L = \cos \phi = \frac{x (1 + b^2) - 2b^2}{x (1 - b^2)} \label{e:Combley}
\end{equation}
where $b$ represents the mass ratio of the decay product (muon) to the decay parent (pion): 
\begin{equation}
b = \frac{m_\mu}{m_\pi} = 0.757.
\end{equation}
We may use this result to estimate the polarization, and hence the spread in the spin angles for comparison with our earlier results.  The pions selected from the target station have a typical momentum of approximately 3.09 GeV/$c$.  Likewise, upon decay, the resulting muon beam will have a similar though slightly lower momentum.  The spread in momenta, however, is dictated by the momentum acceptance of the beam line system.   For the Muon g-2 beam delivery system the momentum acceptance is approximately $\pm 2$\% or so.  Thus, the accepted muons should have a momentum spread on the order of $\pm 2$\% of the average momentum of the pion beam.  Hence, for our case, we can assume a typical value of $x \approx 0.98$.  We can see the results of a plot of Eq.~\ref{e:Combley} in Fig.~\ref{fig:phi_decays}; $x = 98 \%$  will have an approximate spread of spins of about 0.2 ($\pm 0.1$) rad.  This result is also born out through tracking simulations using G4beamline that include particle decays.

\subsection{Simulations}
The simulations used five sets of random misalignments for the magnets on the M2M3 delivery system and the Delivery Ring, varying from 100 to 500 $\mu$m for the rms quadrupole magnet misalignment.  Particle decays are present in each case.  The rms spread of the spin (in radians) is plotted against the rms of the displacement for the $x$ direction.  Similar results are obtained for displacements in $y$ as well as for combinations of both.  

\begin{figure}
\centering
\includegraphics[height=3.5in]{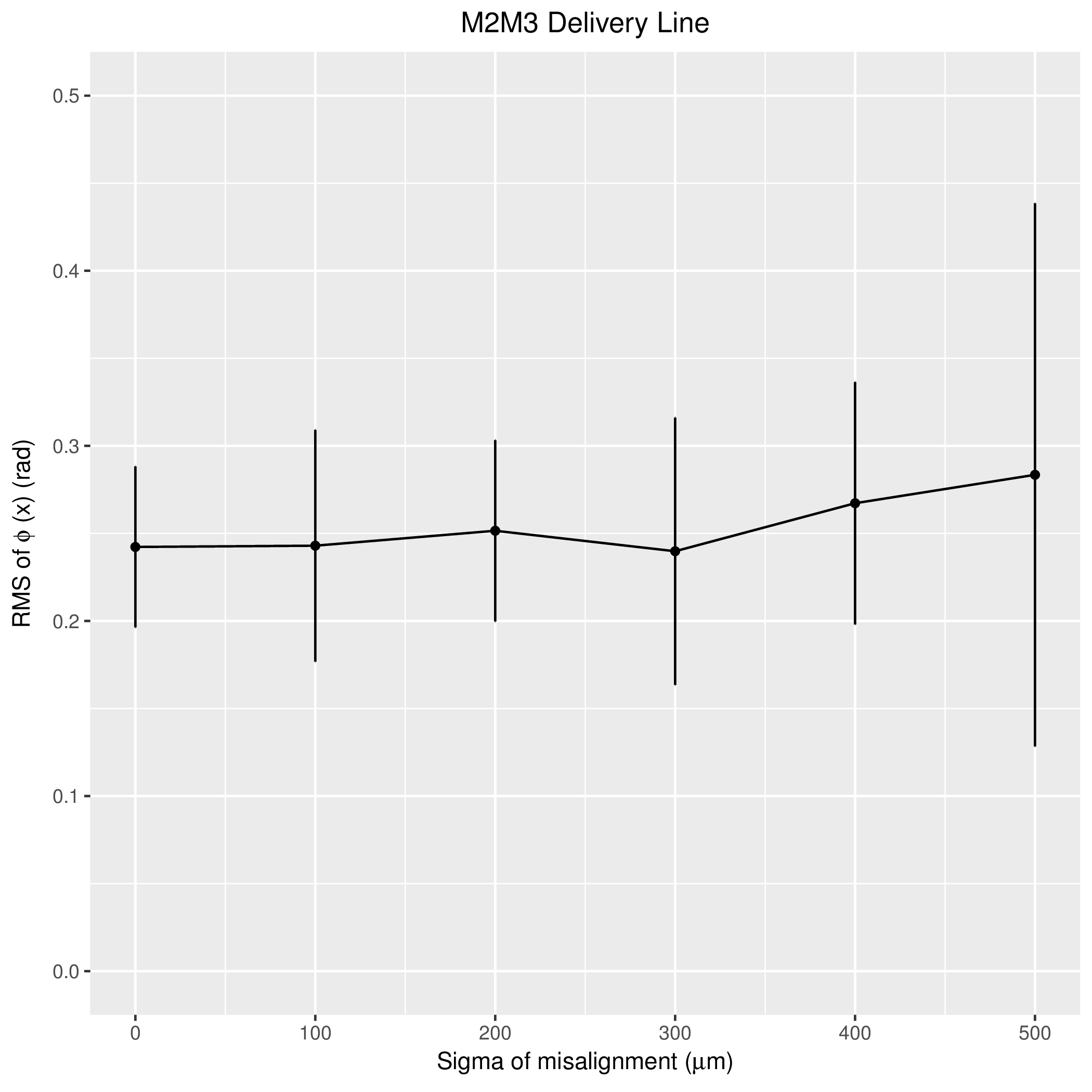} 
\caption[M2M3 Polarization in X direction]{Spin spread in the $x$ direction at end of M2M3 Line.}
\label{fig:M2M3_X}
\end{figure}
\begin{figure}
\centering
\includegraphics[height=3.5in]{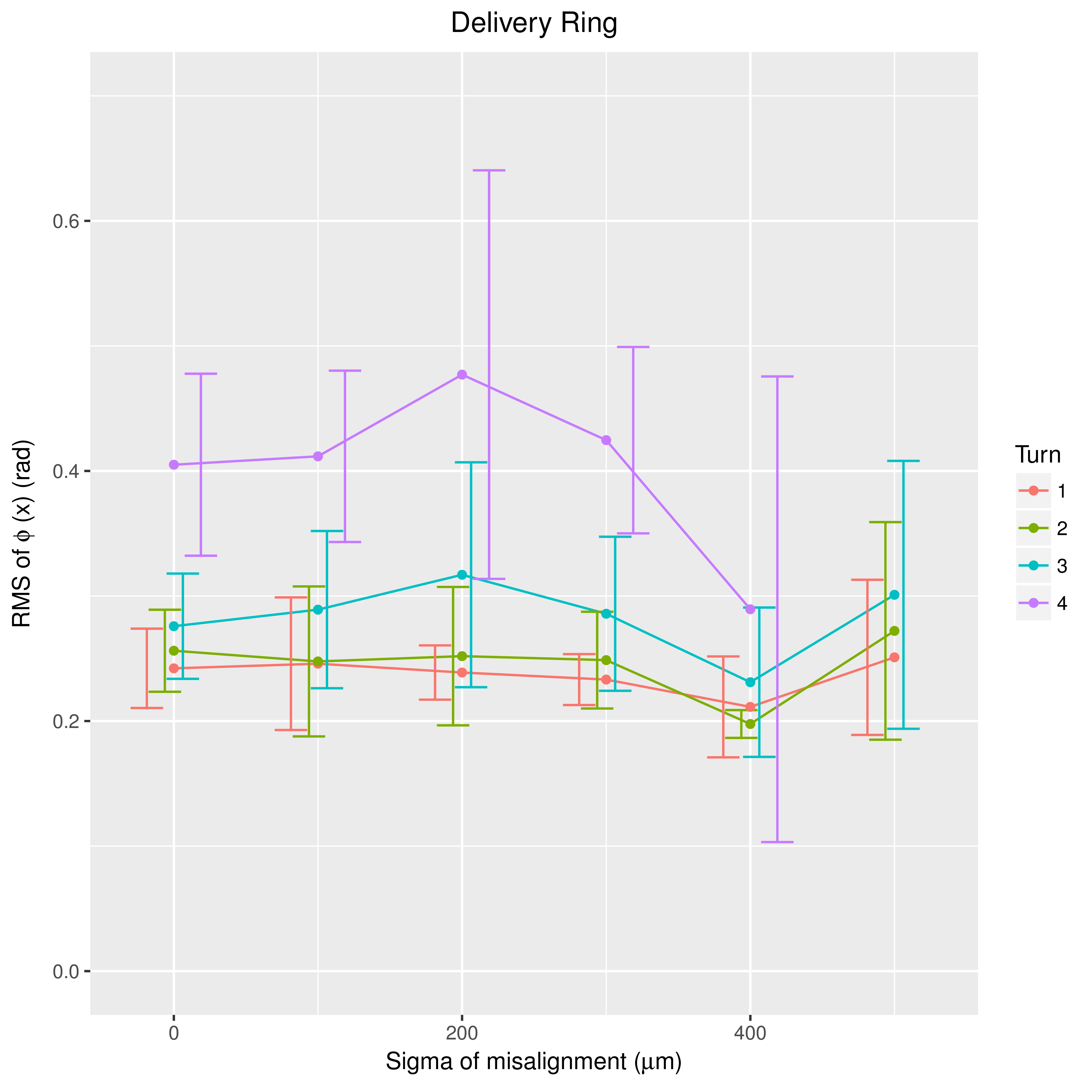} 
\caption[DR Polarization in X direction]{Spin spread in the $x$ direction at end of each turn in the delivery ring. The error bars are associated with the nearest same-color data points.}
\label{fig:DR_X}
\end{figure}

As is shown Figs.~\ref{fig:M2M3_X} and \ref{fig:DR_X}, there exists very little effect for the magnitude of the displacements used in the simulations on the rms spread in spin. 
At most, the displacement effects are seen to be on the order of 0.25 mrad for a displacement of $250~\mu$m.   The magnitude ($\pm$ 100~mrad) of the final rms spin spread from the simulations is in line with estimates due to particle decays.   For Fig.~\ref{fig:DR_X}, the increase in rms spin spread is evident as the number of revolutions (turns) about the Delivery Ring increases from one to four, both due to the additional rotations caused by the momentum spread and by the effects of passing through the misaligned magnets a greater number of times.  The natural polarization coming from pion to muon decay is still the predominant effect.

\section{Concluding Remarks}

\begin{table}
\label{tab:Results}
\begin{center}
\begin{tabular}{|c|c|} \hline
Source & $\phi_{rms}$\\ \hline \hline 
Emittance & $\approx 2.7$~mrad\\ \hline 
Misalignments & $\approx 1$~mrad\\ \hline
Momentum spread & $\approx 11$~mrad \\ \hline
Particle decays & $\approx 200~(\pm 100)$~mrad \\
\hline 
\end{tabular}
\caption[Summary of Estimates]{The table illustrates the magnitude of the effects on the rms of the spin spread due to the individual factors.}
\end{center}
\end{table}

In table \ref{tab:Results}, we can see a comparison of the magnitude of the effects on the rms spread in the spin orientations of a particle beam. For the purposes of the $g$-2 experiment at Fermilab,  the resulting final polarization overwhelmingly is due to the decay process that produces muons from pions.  The other effects studied are small in comparison.

More generally speaking, however, two important correlations are noted.  First, a correlation between momentum and final polarization exists since higher momentum particles precess more in a magnetic field than do lower momentum particles.  Second, there will be a correlation between the amplitudes of betatron oscillations and the spin direction, since particles further away from the ideal path experience a stronger corrective focusing in a quadrupole field and thus a greater precession. These two results will be important in the final analysis of the Muon g-2 experimental results.  Additionally, other particles such as protons and ions have much larger anomalous magnetic moments and hence will have greater precession in a magnetic field, and so the insights gained here will be more important in future research focusing on highly polarized hadron beams for precision physics, such as in EDM searches for example.

\bigskip
The authors would like to acknowledge the assistance of Diktys Stratakis and Brian Drendel. \\

This work was supported by the National Science Foundation Grant 1623691.

\end{document}

%% file: econfmacros.tex
\def\beq{\begin{equation}}
\def\eeq#1{\label{#1}\end{equation}}
\def\eeqn{\end{equation}}

\def\beqa{\begin{eqnarray}}
\def\eeqa#1{\label{#1}\end{eqnarray}}
\def\eeqan{\end{eqnarray}}

\let\bar=\overbar

\def\Dslash{\not{\hbox{\kern-4pt $D$}}}
\def\dslash{\not{\hbox{\kern-2pt $\del$}}}

\def\msb{{\bar{\ssstyle M \kern -1pt S}}}

